\documentclass[twocolumn,showpacs,preprintnumbers,amsmath,amssymb]{revtex4}
\usepackage{graphicx}
\usepackage{dcolumn}
\usepackage{bm}

\begin{document}
\preprint{SPhT-t04/105; ECT*-04-25}
\title{Non Perturbative Renormalization Group, momentum dependence of $n$-point functions and the transition temperature of the weakly interacting Bose gas}

\author{Jean-Paul Blaizot}
   \email{blaizot@spht.saclay.cea.fr}
   \affiliation{Service de Physique Th\'eorique, CEA/DSM/SPhT,\\ 
Unit\'e de recherche associ\'ee au CNRS\\
   91191 Gif-sur-Yvette Cedex, France.}
   \author{Ram\'on M\'endez Galain}
   \email{mendezg@fing.edu.uy} \author{Nicol\'as Wschebor}\email{nicws@fing.edu.uy}
   \affiliation{Instituto de F\'{\i}sica, Facultad de Ingenier\'{\i}a, 
J.H.y Reissig 565, 11000
   Montevideo, Uruguay}

   \date{\today}
\begin{abstract}

We propose a new approximation scheme to solve the Non Perturbative
Renormalization Group equations and obtain  the
full momentum dependence of  $n$-point functions. This scheme 
involves an iteration procedure built
on an extension of the Local Potential Approximation commonly 
used within the Non Perturbative 
Renormalization Group. Perturbative and scaling regimes are accurately reproduced. The method is applied to the
calculation of the shift
$\Delta T_c$ in the transition temperature of the weakly repulsive Bose gas, a quantity which is very sensitive to all momenta intermediate between these two
regions. The  leading order result is in  agreement with  lattice calculations,
  albeit with a theoretical uncertainty of about 25\%. The  
 next-to-leading order differs by about 10\% from the best accepted result. 
\end{abstract}

\maketitle

\newcommand \beq{\begin{eqnarray}}
\newcommand \eeq{\end{eqnarray}}
\newcommand{\ba}{\begin{eqnarray}}
\newcommand{\ea}{\end{eqnarray}}\def\psib{\psi}
\def\phib{\phi}
\def\r{{\rm r}}
\def\d{{\rm d}}

\input epsf


\def\square{\hbox{{$\sqcup$}\llap{$\sqcap$}}}
\def\grad{\nabla}
\def\del{\partial}

\def\frac#1#2{{#1 \over #2}}
\def\smallfrac#1#2{{\scriptstyle {#1 \over #2}}}
\def\half{\ifinner {\scriptstyle {1 \over 2}}
    \else {1 \over 2} \fi}


\def\bra#1{\langle#1\vert}
\def\ket#1{\vert#1\rangle}


\def\simge{\mathrel{%
    \rlap{\raise 0.511ex \hbox{$>$}}{\lower 0.511ex \hbox{$\sim$}}}}
\def\simle{\mathrel{
    \rlap{\raise 0.511ex \hbox{$<$}}{\lower 0.511ex \hbox{$\sim$}}}}


\def\buildchar#1#2#3{{\null\!
    \mathop#1\limits^{#2}_{#3}
    \!\null}}
\def\overcirc#1{\buildchar{#1}{\circ}{}}


\def\slashchar#1{\setbox0=\hbox{$#1$}
    \dimen0=\wd0
    \setbox1=\hbox{/} \dimen1=\wd1
    \ifdim\dimen0>\dimen1
       \rlap{\hbox to \dimen0{\hfil/\hfil}}
       #1
    \else
       \rlap{\hbox to \dimen1{\hfil$#1$\hfil}}
       /
    \fi}


\def\real{\mathop{\rm Re}\nolimits}     
\def\imag{\mathop{\rm Im}\nolimits}     

\def\tr{\mathop{\rm tr}\nolimits}       
\def\Tr{\mathop{\rm Tr}\nolimits}       
\def\Det{\mathop{\rm Det}\nolimits}     

\def\mod{\mathop{\rm mod}\nolimits}     
\def\wrt{\mathop{\rm wrt}\nolimits}     


\def\TeV{{\rm TeV}}                     
\def\GeV{{\rm GeV}}                     
\def\MeV{{\rm MeV}}                     
\def\KeV{{\rm KeV}}                     
\def\eV{{\rm eV}}                       

\def\mb{{\rm mb}}                       
\def\mub{\hbox{$\mu$b}}                 
\def\nb{{\rm nb}}                       
\def\pb{{\rm pb}}                       

%
%

\def\picture #1 by #2 (#3){
   \vbox to #2{
     \hrule width #1 height 0pt depth 0pt
     \vfill
     \special{picture #3} 
     }
   }

\def\scaledpicture #1 by #2 (#3 scaled #4){{
   \dimen0=#1 \dimen1=#2
   \divide\dimen0 by 1000 \multiply\dimen0 by #4
   \divide\dimen1 by 1000 \multiply\dimen1 by #4
   \picture \dimen0 by \dimen1 (#3 scaled #4)}
   }

\def\centerpicture #1 by #2 (#3 scaled #4){
    \dimen0=#1 \dimen1=#2
     \divide\dimen0 by 1000 \multiply\dimen0 by #4
     \divide\dimen1 by 1000 \multiply\dimen1 by #4
          \noindent
          \vbox{
             \hspace*{\fill}
             \picture \dimen0 by \dimen1 (#3 scaled #4)
             \hspace*{\fill}
             \vfill}}


\def\figfermass{\centerpicture 122.4mm by 32.46mm
  (fermass scaled 750)}

%

In this paper we present a new and 
systematic method of approximation
for the Non Perturbative Renormalisation Group (NPRG)\cite{Wetterich93,Ellwanger93,Morris94},
which allows one to get, in a simple way, the full momentum 
dependence of the $n$-point
functions. During the last years, the NPRG has been applied successfully to a variety of physical problems (for  reviews, see e.g. \cite{Bagnuls:2000ae,Berges02}).
In most cases, the solution of the NPRG equations involves a derivative expansion which only allows for the determination of the $n$-point functions and their
derivatives at vanishing external momenta.  There are however situations where the knowledge of the full momentum dependence of the $n$-point functions is needed. A
simple example is that of the calculation of the shift $\Delta T_c$ of the transition temperature of a weakly  interacting Bose gas
\cite{bigbec}. This will be used here as a test of our new approximation scheme.

In order to present our method, we consider a scalar $\phib^4$ theory in $d$ dimension with $O(N)$ symmetry:
\beq
{\cal S} = \int \left\lbrace{ 1 \over 2} \left[
\nabla \phib (x) \right]^2+{1 \over 2}r
\phib^2 (x)+{u \over 4!} \left[ \phib^2(x) \right]^2 \right\rbrace \d^{d}x\,.
\label{eactON}
\eeq
The field
$\phib(x)$ has $N$ real components $\phi_i(x)$, with $i=1,\cdots,N$.

  The starting point of the NPRG is a modification of the classical 
action (\ref{eactON}), to
which is added
 \beq
  \Delta S_\kappa[\phi] =\frac{1}{2} \int \frac{{\rm d}^dp}{(2\pi)^d} 
R_\kappa(p)
\phi_i(p)\phi_i(-p).
  \eeq\normalsize
The role of $\Delta S_\kappa$ is to suppress the fluctuations with 
momenta $p\simle \kappa$, while leaving unaffected the modes with
$p\simge \kappa$. Thus, typically
$
R_{\kappa}(p)\to\kappa^2$ when $ p \ll \kappa$, and $R_{\kappa}(p)\to 0$ 
 when $ p\simge \kappa$.
There is a large
freedom in the choice of $R_\kappa(p)$, abundantly discussed in the literature 
\cite{Ball95,Comellas98,Litim,Canet02}. We have used a cut-off function
proposed by Litim
\cite{Litim}: $R_\kappa(p)\propto(\kappa^2-p^2)\,
\theta(\kappa^2-p^2)$, which allows many calculations to be done analytically.

The  NPRG equations  can be written as an infinite hierarchy of flow 
equations describing how the
$n-$point functions evolve with the infrared cut-off 
$\kappa$ \cite{Wetterich93,Ellwanger93,Morris94}.
Here, we write explicitly only the equation for the self-energy
$\Sigma$ and that for the 4-point function $\Gamma^{(4)}$, which is 
enough for the  purpose of explaining our approximation
scheme. For vanishing fields, $\langle\phi\rangle=0$,  the flow equation for 
$\Sigma_{ij}(\kappa;p)\equiv \delta_{ij} \Sigma(\kappa;p)$
reads
\small
\begin{eqnarray}\label{2pointfct}
&&\hspace{-1cm}\delta_{ij}\partial_\kappa \Sigma(\kappa;p)=\nonumber\\
&&-\frac{1}{2}\int \frac{d^d q}{(2\pi)^d} \partial_\kappa 
R_\kappa(q)G_\kappa^2(q)
\Gamma^{(4)}_{ijll}(\kappa;p,-p,q,-q),
\eeq\normalsize and that for 
$\Gamma^{(4)}_{ijkl}(\kappa;p_1,p_2,p_3,p_4)$ is given by
\small
\begin{eqnarray}\label{4pointfct}
&&\hspace{-0.5cm}\partial_\kappa \Gamma^{(4)}_{ijkl}(\kappa;p_1,p_2,p_3,p_4)=
\int \frac{d^d q}{(2\pi)^d} \partial_\kappa R_\kappa(q)G^2(\kappa;q) 
\nonumber \\
&&\times\left\{\Gamma^{(4)}_{ijmn}(\kappa;p_1,p_2,q,-q-p_1-p_2)G(\kappa;q+p_1+p_2) 
\right.\nonumber \\
&&\left.\times\Gamma^{(4)}_{klnm}(\kappa;p_3,p_4,q-p_3-p_4,-q)
+ {\rm permutations} \right.\nonumber \\
&&-\frac{1}{2}\left.
\Gamma^{(6)}_{ijklmm}(\kappa;p_1,p_2,p_3,p_4,q,-q)\right\}.
\end{eqnarray}
\normalsize
More generally, the equations for $\Gamma^{(n)}$ involves all the $\Gamma^{(m)}$ with $m\le n\le n+2$.
In  these equations, $G^{-1}(\kappa;p)=p^2+R_\kappa(p)+\Sigma(\kappa,p)$.
These equations are to be solved with the boundary condition that as 
$\kappa\to \Lambda$, with $\Lambda$ the microscopic scale,  the
  $n$-point functions take their classical values, read on the action 
(\ref{eactON}). In particular,
$\Gamma^{(4)}_{ijkl}(\Lambda;p_1,p_2,p_3,p_4)=
(u/3)(\delta_{ij}\delta_{kl}+\delta_{ik}\delta_{jl}+\delta_{il}\delta_{jk}) 
$. As $\kappa\to 0$ the 
$n-$point functions go to their physical values.

Clearly, in general, the NPRG  hierarchy of equations can only be solved approximately. But a virtue of  the NPRG is precisely to suggest 
 approximations which are not easily formulated in other, more conventional, approaches.  A popular one 
is the derivative expansion 
\cite{Wetterich93,Morris94b}. In lowest order it consists in
ignoring  the momentum
dependence of vertices in the right hand side of the flow equation, as well as  field renormalization. In this 
approximation, usually referred to as
the Local Potential Approximation (LPA),    the hierarchy collapses 
into a closed
equation for the effective potential {\bf $V_\kappa(\phi^2)$}. An 
interesting  improvement of the LPA, which we
refer to as the LPA',    takes into account   a running field 
renormalisation constant
$Z_\kappa$ and allows for a non trivial anomalous dimension, determined from the cut-off dependence of $Z_\kappa$, 
$\eta=-\kappa\del_\kappa\ln
Z_\kappa\,$ \cite{Wetterich93}. The  solution of the LPA' is well documented in the
literature (see e.g. \cite{Berges02,Canet02}). It will be used as an input in our approximation scheme. Let us emphasize that for massless theories, the derivative
expansion is an expansion in terms of $p^2/\kappa^2$ and  therefore, in the physical limit $\kappa\to 0$, it makes sense only for vanishing external momenta,
$p=0$.  Higher orders in the derivative expansion
\cite{Canet02} may  improve the accuracy of the LPA but, again, this concerns only the $n$-point functions or their derivatives at zero momenta. 

In order to get the full momentum dependence of the $n$-point functions, we propose here to solve the hierarchy  of the NPRG equations through an iteration
procedure. The iteration of the  NPRG equations starting with the classical values of the
$n$-point functions as  initial input reconstructs the usual
  loop expansion --- see e.g. \cite{Polchinski,Bonini}. Thus one expects generically the iterations of the exact flow equations to correctly 
account for the high momentum behaviour. This is not so however for the low momentum region. There,  our approximation scheme will exploit the fact that the
LPA' is a good approximation in the limit of vanishing momenta.  

The iteration procedure
starts with an initial guess for the $n$-point functions to be used in the right hand side of the flow equations.  Integrating the flow equation of a given
$n$-point function gives then the {\it leading order} (LO) estimate for that $n$-point function. Inserting the leading order of the $n$-point functions thus
obtained in the right hand side of the flow equations and integrating gives then the {\it next-to-leading order} (NLO) estimate of the $n$-point functions. And so
on. 

There is no small parameter controling the convergence of the process, and the terminology LO, NLO, etc. that we just used  refers merely to the
number of iterations involved in the calculation of the $n$-point function considered.  
Obviously, the calculations become increasingly complicated as the number of iterations
increases, and it is essential that the initial guess be as close as possible to the exact
solution so that only one or two iterations suffice to get an accurate result. The main effort focuses then on 
 the construction of such a good initial guess for the solution, to which we now turn.

We shall be guided
by a crucial property of the NPRG, that of decoupling of the various 
momentum scales:
  for given external momenta $p\simle \kappa$,
the flow of the $n$-point functions is dominated by internal momenta
$q \simle \kappa$ (the derivative $\del_\kappa
R_\kappa(q)$ limits the range of integration in the flow equations to 
$q\simle \kappa$), and
  when
$\kappa$ becomes lower than
$p$ the flow essentially stops (the external momenta playing the role 
of infrared regulators).

Consider first the 2-point function. A reasonable initial guess for the propagator to be used in the r.h.s.  of the flow equations when calculating the leading
order estimate of the $n$-point functions is the   LPA' propagator 
$G^{-1}_{LPA'}(\kappa;p)=Z_\kappa p^2
+m^2_\kappa +R_\kappa(p)$, where 
$m^2_\kappa$ is the running LPA' mass:  it is a good approximation when $p\simle\kappa$, and for $p\simge\kappa$, $G^{-1}(\kappa;p)$ goes quickly to zero and its
precise form does not matter. 

A good initial ansatz for the other $n$-point functions  (with $n>2$) will be obtained by solving
the flow equations for the  $\Gamma^{(n)}$'s, with the following three approximations.

 Our first approximation (A1) assumes that for  $q\simle
\kappa$, and any set of external momenta $\lbrace 
p_1,p_2,...,p_n\rbrace$ we have
\small\beq
\left|\frac{\Gamma^{(n)}(\kappa;p_1,...,p_{n-1}+q,p_n-q)-\Gamma^{(n)}(\kappa;p_1,...,p_n)}
{\Gamma^{(n)}(\kappa;p_1,...,p_n)}\right|. \ll 1\nonumber\\
\eeq\normalsize
This is certainly true for momenta $\lbrace 
p_1,p_2,...,p_n\rbrace$ whose norms are much larger than
$\kappa$, if  $\Gamma^{(n)}$ is a smooth function of its arguments. 
Similarly, for vanishing momenta, and assuming
again that $\Gamma^{(n)}$ is a smooth function, the condition above 
is equivalent to saying that one can expand
in powers of $q^2/\kappa^2$, which leads in zeroth order to the  LPA,   a 
good approximation. Based on this assumption, we shall set $q=0$ in 
all $\Gamma^{(n)}$ ($n>2$), and
factor them out of the integral in the r.h.s. of the flow equations.

\begin{figure}[h!]
\includegraphics[scale=.35,angle=-90]{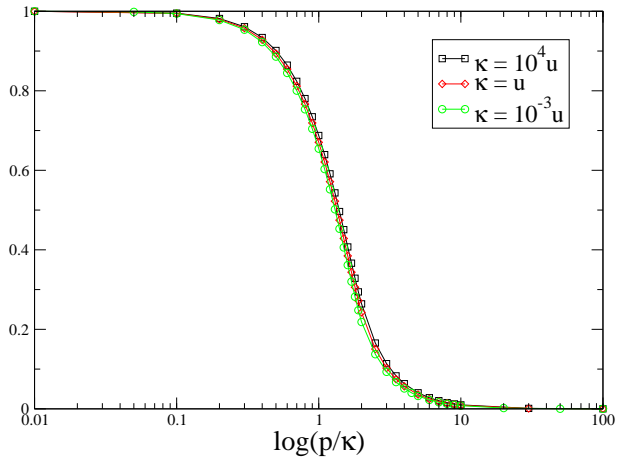}
\caption{The function $I_\kappa(p)/I_\kappa(0)$ as a function of 
$\ln(p/\kappa)$\label{fig:theta} calculated with the leading order propagator at criticality}
\end{figure}

The second approximation (A2) concerns the propagators in the flow equation, for which we make the replacements:
\beq\label{A2}
G(p+q)\longrightarrow G_{LPA'}(q)\,\Theta(1-\frac{\alpha^2 p^2}{\kappa^2})
\eeq
where $\alpha$ is an adjustable parameter. As an illustration of the quality of this approximation, we show in Fig.~\ref{fig:theta} the ratio 
$I_\kappa(p)/I_\kappa(0)$ where 
\beq
I_\kappa(p)\equiv\int \frac{d^d q}{(2\pi)^d} \partial_\kappa 
R_\kappa(q)G^{2}(\kappa;q)G(\kappa;p+q)  
\eeq\normalsize
is the integral wich remains in Eq.~(\ref{4pointfct}) after approximation A1. 
This ratio, as a function of $p^2/\kappa^2$,  looks indeed like a 
 step function, with a weak residual $\kappa$ dependence. 
 The approximation A2 amounts to truncate severely the high momentum
tails of the propagators. This causes a slight inaccuracy at high momenta, and a dependence of the leading order 
results on the value of $\alpha$. One may fix  
$\alpha$ so that the inflexion point of the curve in Fig.~\ref{fig:theta} is at 
$\alpha p=\kappa$. One then obtains typically 
$\alpha\approx .9$. One can also adjust $\alpha$ so that the integral
over $\kappa$ of $I_\kappa(p)$ is identical to that  
of $I_\kappa (0)\Theta(\kappa^2-\alpha^2p^2)$.
This   yields typically  
$\alpha\approx .6$. We  regard these two possible choices as extremes and adopt the value
$\alpha=.75\pm .15$ for our  leading order estimate.

  The third approximation (A3) concerns the function $\Gamma^{(n+2)}$ in the equation for $\Gamma^{(n)}$. For this we use an ansatz inspired by the 
expressions of the various $n$-point functions in the large $N$ limit \cite{BMW03}. To be specific, consider 
$\Gamma^{(6)}$ in the equation for $\Gamma^{(4)}$. The approximation A3, combined  with A1 and A2, leads to the result that  
the contribution of $\Gamma^{(6)}$ to the 
r.h.s. of Eq.~(\ref{4pointfct}) is proportional to 
 that of  the other terms, the proportionality coefficient $F_\kappa$ being only a function 
of $\kappa$. The same proportionality also holds in
the LPA regime, which allows us to use the LPA to determine 
$F_\kappa$.

Approximations A1-A3, when applied to Eq.~(\ref{4pointfct}), yield  the following equation for our initial 
guess for $\Gamma^{(4)}$:
\small\beq
\label{gamma40new}
&&\hspace{-0.6cm}\partial_\kappa\Gamma^{(4)}_{ijkl}(\kappa;p_1,p_2,p_3,p_4)=I^{(3)}_\kappa(0)\, 
(1-F_\kappa)
  \nonumber \\
&&\hspace{.1cm}\times\left\{ \Theta\!\!\left(\kappa^2-\alpha^2 
{(p_1+p_2)^2})\right)
\Gamma^{(4)}_{ijmn}(\kappa;p_1,p_2,0,-p_1-p_2)\right. \nonumber\\
&&\hspace{1cm}\left.\times\Gamma^{(4)}_{klnm}(\kappa;p_3,p_4,-p_3-p_4,0) 
+{\rm permutations}
\right\}.\nonumber\\
\eeq\normalsize
This equation  can be solved analytically in 
terms of the solution of the LPA'\cite{BMW03}. This is done by steps,
starting form the momentum domain
$\alpha^2(p_1+p_2)^2,\alpha^2(p_1+p_3)^2,\alpha^2(p_1+p_4)^2\le 
\kappa^2$, where it can be verified that the solution is  that of the LPA' itself. The 
solution can be written explicitely  in the various momentum regions defined by
the $\Theta$-functions occuring in Eq.~(\ref{gamma40new}).

\begin{figure}
\includegraphics[scale=.35,angle=-90]{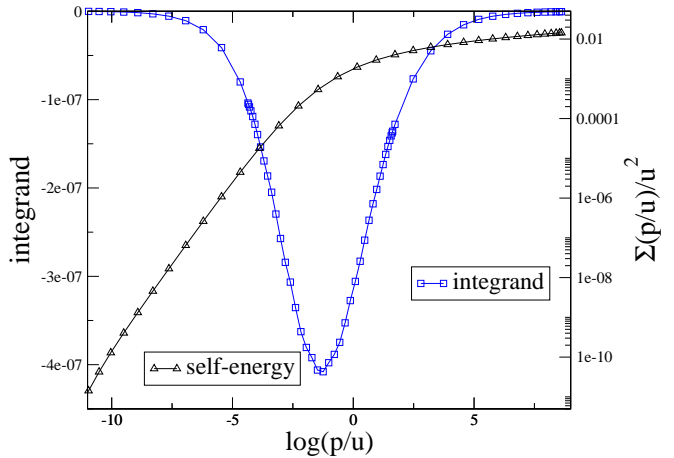}
\caption{The leading order function $\sigma(x=p/u)$ at criticality,  and the integrand of Eq. \ref{integralc} as a function of 
$\ln(p/u)$, for $u\simle 10^{-4}$ for which $\sigma(x)$ is independent of $u$\label{fig:sigma} \cite{club}}
\end{figure}

The leading order result for the self-energy is obtained by using the solution of 
Eq.~(\ref{gamma40new}) together with $G_{LPA'}$ in the r.h.s. of  Eq.~(\ref{2pointfct}) and integrating 
numerically over $\kappa$. The resulting self-energy at criticality, i.e., for the value of $r$ that yields a vanishing physical mass, is shown in 
Fig.~\ref{fig:sigma}  for $N=2$ and $d=3$. It has good behaviour at both   low and high momenta, independently of the value of $\alpha$: In the scaling regime, one recovers the
expected result
$p^2 +\Sigma(p) \propto p^{2-\eta}$, where  the anomalous dimension is
$\eta\simeq 0.043$; this is only slightly larger than values 
determined by the most accurate
available methods  \cite{guida}. One can show  \cite{BMW03} that the value of
$\eta$ coincides with that obtained within the LPA', but now it results directly from the
momentum dependence of the self-energy, rather than the cut-off dependence of the field
renormalization. At large momenta, the logarithmic behaviour given by
perturbation theory  is  reproduced. However,   the coefficient of the logarithm differs
by about 10\%  from the expected one. This problem, cured in NLO, finds its origin in  the
approximation A2 which truncates the high momentum tails of the
propagators.

We have applied this approximation scheme to the calculation   of 
 the shift $\Delta T_c$ of the transition temperature of a weakly 
interacting Bose gas. 
It has been shown recently that
$\Delta T_c$ is
linear in  $an^{1/3}$ \cite{club}, where $a$ is the scattering length and $n$ the particle density:
\beq\label{deltaTc}
\frac{\Delta T_c}{T_c^0}=c \,\,a n^{1/3}.
\eeq 
Here $T_c^0$ is the  condensation temperature of the ideal gas, given by $ n\lambda_c^3=\zeta(3/2)$
with $\lambda_c^2=2\pi/mT_c^0$ ($m$ is the boson mass), and $\Delta T_c=T_c-T_c^0$ with
$T_c$  the transition temperature of the interacting system. As shown in Ref.~\cite{club},
 the coefficient $c$ can be related to the change $\Delta\langle \phi^2\rangle$ of the fluctuation 
of the field described 
by the action (\ref{eactON}):   $
c\,=-256 \pi^3 \left(\zeta(3/2)\right)^{-4/3} \, 
({\Delta\langle\phi^2\rangle}/{Nu})$, with $N=2$. The parameters $r$ and $u$ in
 (\ref{eactON}) are then related to the scattering length $a$ and the chemical 
potential $\mu$ by:
$u=96 \pi^2 {a}/{\lambda^2}$, and $r=-2mT\mu$ \cite{BigN}. 

The best numerical estimates for $\Delta\langle \phi^2\rangle$, and hence for $c$,  are 
those which have been obtained using the lattice technique by two groups, with the 
results: $c=1.32\pm0.02$
\cite{arnold} and $c=1.29\pm 0.05$ \cite{svistonov}. The availabilty 
of these results has turned the
calculation of $c$ into a testing ground for other non perturbative methods: 
expansion in  $1/N$
\cite{BigN,Arnold:2000ef}, optimized perturbation theory \cite{souza},
  resummed    perturbative  calculations to high loop orders 
\cite{Kastening:2003iu}. Note that while the
latter methods yield critical exponents with several significant 
digits, they predict $c$ with only a 10\%
accuracy. This illustrates the difficulty of getting an accurate 
determination of $c$ using  (semi) analytical
techniques. 

To understand better the origin of the difficulty, let us write $\Delta\langle \phi^2\rangle$ as the following integral \cite{club}
\beq\label{integralc}
\frac{\Delta\langle \phi_i^2\rangle}{Nu}=- \int\frac{{\rm }dx}{2\pi^2}\,
\frac{\sigma(x)}{x^2+\sigma(x)},
\eeq
where $\sigma(x)=u^{-2}\Sigma(p=xu)$, with $\Sigma(p)$  the self-energy at  criticality, i.e., $\Sigma(0)=0$.  The integrand of Eq.~(\ref{integralc}), at leading order, is shown in Fig.~2.
The momentum at the minimum 
defines the typical scale which separates the scaling region  from  the high 
momentum region where perturbation theory applies. 
The difficulty in getting a precise evaluation of the integral 
(\ref{integralc}) is
that it requires an  accurate determination of $\Sigma(p)$ in a large 
region of momenta including the crossover region between two 
different physical regimes
\cite{bigbec,BigN}.

In leading order, we obtain $c\approx  1.3\pm .3$, in agreement with lattice results \cite{Ledowski03}. The large uncertainty
comes from the freedom in the choice of
$\alpha$ that we have discussed above.  We have also studied
 the NLO \cite{BMW03}. This involves the leading order for $\Gamma^{(4)}$,
which requires the initial guess for $\Gamma^{(6)}$. The latter is obtained by following  the same strategy as for
$\Gamma^{(4)}$. The  NLO result for $c$ still depends 
on $\alpha$, however in such a way that   one can
 fix $\alpha$ from a criterium of  fastest apparent 
convergence: in fact there is a value of $\alpha$ for which the correction to the LO result vanishes.
This value is $\alpha=.83$ leading to $c=1.44$. 

Results also exist for other values of $N$, from lattice calcultions \cite{arnold,svistonov} or
perturbation theory to 7-loops
\cite{Kastening:2003iu}. For
$N=1$, we get
$c=1.20$, to  be compared with
$c=1.09\pm 0.09$ (lattice) or $1.07\pm 0.10$ (7-loops); for $N=3$, we 
get $c=1.62$, to be compared with 
$1.43\pm 0.11$ (7-loops); for $N=4$, we 
get $c=1.79$, to be compared with $1.60\pm .10$ (lattice) and 
$1.54\pm 0.11$ (7-loops). Our results thus differ by 10 to 15\% from the best
estimates. In its present numerical implementation, our method looses accuracy in the  large $N$ limit,  where we get $c\approx 2.8$ instead of the exact
result $c=2.3$
\cite{BigN}.

In summary, we have proposed a simple method to get the momentum 
dependence of the $n$-point functions using the NPRG. Using well motivated approximations we have 
produced a
  simple solution which provides an excellent starting point for a more accurate iterative solution, as revealed by its
application to the calculation of $\Delta T_c$. Although  there is no small  parameter that controls the convergence of the iterative procedure,  
the NLO calculation suggests that the scheme is stable.


\begin{thebibliography}{99}

\bibitem{Wetterich93}  C.Wetterich, Phys. Lett., {\bf B301}, 90 (1993).

\bibitem{Ellwanger93}  U.Ellwanger, Z.Phys., {\bf C58}, 619 (1993).

\bibitem{Morris94}  T.R.Morris, Int. J. Mod. Phys., {\bf A9}, 2411 (1994).


\bibitem{Bagnuls:2000ae}
C.~Bagnuls and C.~Bervillier,
Phys.\ Rept.\  {\bf 348}, 91 (2001).


\bibitem{Berges02}  J.Berges, N.Tetradis, C.Wetterich, Phys. Rept., 
{\bf 363}, 223 (2002).

\bibitem{bigbec} G. Baym, J.-P.  Blaizot, M. Holzmann, F. Lalo\"e, and D.
Vautherin, Eur. Phys. J.  {\bf B24}, 107 (2001).


\bibitem{Ball95} R.D.Ball, P.E.Haagensen, J.I.Latorre and E. Moreno, 
Phys. Lett., {\bf B347}, 80 (1995).

\bibitem{Comellas98} J.Comellas, Nucl. Phys., {\bf B509}, 662 (1998).


\bibitem{Canet02} L.Canet, B.Delamotte, D.Mouhanna and J.Vidal, Phys. 
Rev. {\bf D67} 065004 (2003).

\bibitem{Litim} D.Litim, Phys. Lett., {\bf B486}, 92 (2000); Phys. 
Rev., {\bf D64}, 105007
(2001);  Nucl. Phys., {\bf B631}, 128 (2002); Int.J.Mod.Phys., {\bf 
A16}, 2081 (2001).

\bibitem{Morris94b}  T.R.Morris, Phys. Lett. {\bf B329}, 241 (1994).


\bibitem{Polchinski}
J. Polchinski, Nucl. Phys. {\bf B231}, 269 (1984).

\bibitem{Bonini}
M. Bonini, M. D'Attanasio, G. Marchesini, Nucl.Phys. {\bf B409}, 441 (1993).

\bibitem{BMW03}  J.-P.  Blaizot, R.M\'endez Galain, N. Wschebor, in 
preparation.


\bibitem{guida}
R.~Guida and J. Zinn-justin, J. Phys A{\bf 31}, 8103 (1998).


\bibitem{club} G. Baym, J.-P.  Blaizot, M. Holzmann, F. Lalo\"e, and D.
Vautherin, Phys. Rev. Lett.  {\bf 83}, 1703 (1999).

\bibitem{arnold} P. Arnold and G. Moore,
Phys. Rev. Lett. {\bf 87}, 120401 (2001).

\bibitem{svistonov} V.A.  Kashurnikov, N.~V.  Prokof'ev, and B.~V.
Svistunov, Phys. Rev. Lett. {\bf 87}, 120402 (2001).



\bibitem{BigN} G.\ Baym, J-P.\ Blaizot and J. Zinn-Justin, Europhys.\
Lett.  {\bf 49}, 150 (2000).


\bibitem{Arnold:2000ef}
P.~Arnold and B.~Tomasik,
Phys.\ Rev.\ A {\bf 62}, 063604 (2000).



\bibitem{souza} F. de Souza Cruz, M.B. Pinto, and R.O. Ramos,
Phys. Rev. {\bf B64}, 014515 (2001); Phys.\ Rev.\ A {\bf 65}, 053613 (2002).


\bibitem{Kastening:2003iu}
B.~M.~Kastening,
Phys.\ Rev.\ A {\bf 69}, 043613 (2004).



\bibitem{Ledowski03}  S.Ledowski, N. Hasslmann and P. Kopietz, Phys. Rev. {\bf A 69}, 061601(R)
(2004)  have recently used
the NPRG to calculate the coefficient
$c$, and obtained $c=1.23$.  A critical discussion of their approach  will be
presented in \cite{BMW03}.
We only observe  here that in solving the equation for the
4-point function they ignore terms that we find important.  
\end{thebibliography}
\end{document}